# TESTING DEMAND RESPONSIVE SHARED TRANSPORT SERVICES VIA AGENT-BASED SIMULATIONS

**Giuseppe Inturri, Nadia Giuffrida, Matteo Ignaccolo, Michela Le Pira, Alessandro Pluchino, Andrea Rapisarda**

University of Catania

Abstract

Demand Responsive Shared Transport (DRST) services take advantage of Information and Communication Technologies (ICT), to provide "on demand" transport services booking in real time a ride on a shared vehicle. In this paper, an agent-based model (ABM) is presented to test different the feasibility of different service configurations in a real context. First results show the impact of route choice strategy on the system performance.

**Introduction**

This paper focuses on the potential contribution of innovative Demand Responsive Shared Transport (DRST) services provided by a fleet of vehicles, booked by users via mobile device applications and scheduled in real-time to pick up and drop off passengers in accordance with their needs [1]. The system stands between an expensive conventional exclusive-ride door-to-door service (like a conventional taxi) and a flexible system where a dynamic sharing of trips makes users experiment longer travel distances and times while the vehicles drop off and pick up other passengers (like a conventional transit).

From the operator's point of view, it is important to select the optimal strategy to assign vehicles to passengers' requests, so to perform high load factor and low driven distance (to reduce operation costs), while minimizing the additional time and distances travelers have to experience (to assure the expected level of service).

The city of Ragusa (Italy) is chosen as case study, where an innovative DRST has already been tested. The model is used as a realistic environment where to simulate different scenarios, with simple rules assigned to agents' behavior, in order to explore the transport demand and supply variables that make the DRST service feasible and convenient. The aim is to understand, starting from the micro-interaction between demand and supply agents (i.e. passengers and vehicles), the macroscopic behavior of the system so to monitor, via appropriate indicators, its performance and give suggestions on its planning, management and optimization.

In the last years an increasing attention has been paid on shared transport services. Optimization models have been proposed to solve a dial-a-ride (*Stein, 1978*) or multiple depot vehicle scheduling problems (*Bodin and Golden, 1981*). More recently, simulation models have been developed to study the usability and profitability of dial-a-ride with fixed-route problems (*Shinoda et al, 2004*), the efficient scheduling of dynamic DRT services (*Diana, 2006*), the dynamics of a taxi-sharing system (*D'Orey and Fernandes, 2012*), the effects of using a zoning vs. a no-zoning strategy and time-window settings (*Quadrifoglio et al, 2008*).

Agent-based simulations has proved to be a good technique in this context, also to overcome some limitations linked to a top-down approach. They are suitable to reproduce the interaction among stakeholders involved in transport decision-making (*Le Pira et al, 2018; Marcucci et al, 2017*). They have been proposed to study taxi fleet operations (*Cheng and Nguyen, 2011*), car sharing (*Lopes et al, 2014*) and to investigate DRT services, and developing an open-source simulation testbed (*Čertický et al, 2016*).

Main benefits of ABM are (1) possibility to capture emergent phenomena, (2) providing of a natural description of a system, (3) flexibility, for the easiness to add more entities to the model, to modify behavior, degree of rationality, ability to learn and evolve, and rules of interactions of agents (*Bonabeau, 2002*).

In this paper a new ABM is presented to test the operation of a DRST service under different dispatching configurations. The interaction among vehicles traveling along a semi-flexible route and users walking from their origins to stop nodes to get a transport service to their destinations is simulated. The main novelty of the model relies on the implementation of a GIS-based demand model implying an easy transferability to other contexts. The ABM allows exploring the emergence of optimal operation configuration by identifying *ad-hoc* indicators to monitor the system's performance.

**Description of the model**

An ABM has been built within NetLogo (*Wilensky, 1999*) to test the impact of different vehicle dispatching strategies on the service efficiency and effectiveness.

The model can be described according to its main features, i.e. (i) transport network, (ii) demand model, (iii) agent (user and vehicle) dynamics, (iv) route choice strategies, (v) performance indicators.

*Transport network*. The network consists of a fixed route and three optional routes, composed of network nodes and links, stop nodes and diversion nodes. The network follows the actual road network and a GIS map, reproducing census tracts, is used to implement in the model georeferenced socio-economic data about population, through the GIS extension of NetLogo.

*Demand model*. A user group's (with a maximum prefixed size) trip request is randomly generated with a negative exponential distribution with an average trip rate from an origin $i$ with trip rate proportional to density population, to a destination $j$ with a gravitationally distributed probability.

*Agent (user and vehicle) dynamics*.

If the origin or destination of the trip request is more distant than a prefixed threshold, it assumes the status "rejected". Otherwise, the user group moves to the nearest stop and assumes the status "waiting", until a vehicle (with an appropriate number of available seats) reaches the stop; in this case each user boards and alights at the nearest stop to its required destination, assumes the status of "satisfied" passenger; if the maximum waiting time is overcome, each user gives up and assumes the status of "unsatisfied".

A given number of vehicles, with prefixed seat capacity, generated at random stops, starts traveling along the fixed route at constant speed. At each stop, waiting users are loaded following the First-Come-First-Served queue rule, if the group size is not greater than the available seats. At each diversion node a vehicle can shift to an optional route if waiting users or on-board passengers' destinations are present. The available vehicle seats are updated at each event of passenger loading/unloading.

*Route choice strategies*. All vehicles drive on the fixed route. At diversion nodes a vehicle may drive on a flexible route according with the Route Choice Strategy (RCS) it is assigned to.

In this first version of the model, there are three RCSs:

- "Fully at Random" (FR);
- "All Vehicles drive on All flexible Routes" (AVAR); a prefixed percentage of vehicles can be assigned at random;
- "Each Vehicle is Assigned to a flexible Route" (EVAR); a prefixed percentage of vehicles can be assigned at random.

The randomness component has been considered to add some "noise" to the system, since it has been shown that random strategies can have a beneficial role in increasing the efficiency of social and economic complex systems (Pluchino et al, 2010).

*Performance indicators*. The local strategies determining interaction between passengers and vehicles give raise to global patterns that can be monitored via appropriate performance indicators: total number of passengers transported NP, total driven distance TDD (km), average passenger travel distance APTD (km), average vehicle load factor ALF, passenger travel time in terms of average waiting time AWT (min), average on-board time AoBT (min), average total travel time APTT (min), average vehicle speed AVS (km/h), transport intensity CI (km/pax) as ratio of total driven distance and number of passengers, total user travel time TPTT

(h) (including a penalty of 60 min for each unsatisfied user), vehicle operation cost OC (€) and total unit cost TUC (€/pax), evaluated according to equation (1):

$$TUC\left(\frac{€}{pax}\right) = \frac{TPTT(h) \cdot VOT\left(\frac{€}{h}\right) + OC(€)}{NP(pax)} \qquad (1)$$

Case study

The case study is located at Ragusa, a small-medium city (70,000 inh.) in the south-eastern part of Sicily (Italy), where an innovative DRST service, called MVMANT (http://www.mvmant.com/portfolio-view/ragusa/) has already been experimented in 2016. The city of Ragusa is characterized by two distinct areas, the upper town and the lower and older town of Ragusa Ibla, with a high touristic vocation. MVMANT has connected several park-and-ride facilities with the main destinations in Ragusa Ibla, which is scarcely connected to the center, offering a continuous service with midsize passenger vans.

Fig. 1. shows the fixed (blue) and flexible (orange) routes and census zones colored according to population (from light to dark green).

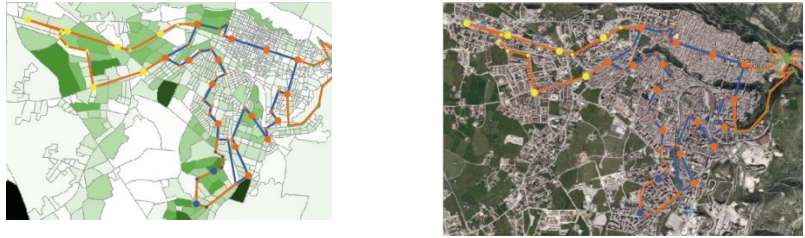

**Fig. 1.** Virtual map (left) and satellite map (right).

The main input variables of the system are:

- service variables, i.e. total simulation time (h), number of vehicles, vehicle maximum capacity (seats), vehicle average speed (km/h);
- demand variables, i.e. demand rate (request/hour), maximum number of passengers per demand, maximum waiting time (min)
- route choice strategy, i.e. FR, EVAR, AVAR, with a variable percentage of randomness.

**Preliminary Results and Conclusions**

For a first test of the model, simulations were performed by considering different route choice strategies (i.e. FR, EVAR, AVAR) with increasing levels of randomness, so to test the overall system performance.

Fig. 2. shows on the left the total hours spent by all passengers (in yellow), while waiting (in light blue), on board (orange) plus a penalty of 60 minutes for each users that waited more than a prefixed threshold at the stop (grey). In dark blue the total number of passengers transported. They decrease with 15 and 30 vehicles because group requests with 3 passengers cannot be satisfied by vehicles with low capacity. On the right, Fig. **2.** shows the total unit cost TUC (€/pax) by the number of vehicles. It is calculated by attributing a monetary value to each hour spent by the passengers in the system (10 €/h), adding the operation cost of vehicles (in the range of 0.5-1.0 €/km according with the vehicle size) and drivers' cost (20 €/h) and dividing the sum by the number of passengers. It can be considered a measure of the total cost sustained by the society (demand and supply) for the mobility of one person. In this case an optimal range of shared services can be identified within a range between 5 and 10 vehicles.

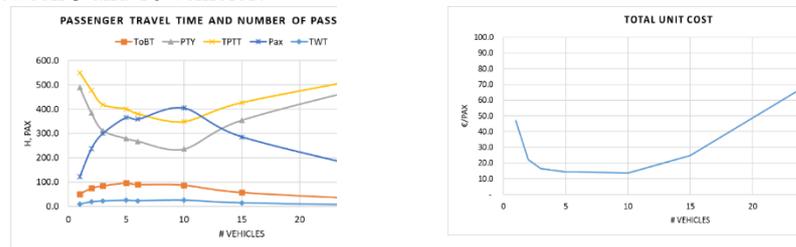

**Fig. 2.** Passenger travel time and number of passengers (on the left) and total unit cost TUC (on the right) (route choice strategy EVAR with 30% randomness; maximum group size=3; total seat capacity=30).

Comparing TUC for EVAR and AVAR strategies with variable randomness (Fig. **3.** ), best results are found with EVAR and no randomness, while AVAR is the worst. This is because the assignment of vehicles to specific routes (EVAR) reduces the empty driven distances. By increasing randomness, it can be seen that the two strategies get closer in terms of TUC. With randomness more than 40% the two strategies are the same with intermediate and almost constant performance. It can be concluded that a certain rate of randomness is beneficial for the AVAR strategy, since it implies that not all vehicles will simultaneously explore all the routes if demand is present (thus reducing the empty driven distance). Vice versa, EVAR strategy works better without randomness.

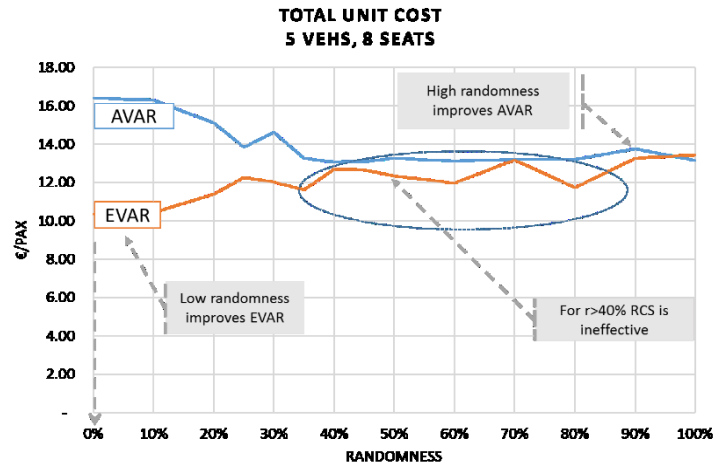

**Fig. 3.** Total unit cost by randomness (number of vehicles= 5; seat capacity=8; maximum group size= 1)

This paper presented an agent-based model able to simulate flexible demand responsive shared transport (DRST) services. Different strategies can be tested by changing the service variables and assigning simple rules to the agents so to explore, starting from the microinteraction between passengers and vehicles, the macroscopic behavior of the system. For a first test, the city of Ragusa (Italy) was chosen as case study, where an innovative DRST has already been experimented and different scenarios have been reproduced by changing fleet composition and vehicle dispatching strategy.

Simulation results show that the service quality and performance considerably vary in relation to the number and capacity of vehicles. In particular, given a fixed supply (in terms of total seat capacity), many vehicles with low capacity decrease the passenger travel time (and cost) and increase the operator costs, while few high-capacity vehicles perform better from an operator's point of view. An optimal range can be found by considering a total unit cost accounting both for passenger travel time and vehicle operation cost.

Besides, different vehicle dispatching strategies have been simulated to test different system configurations (from more flexible to more fixed route strategies). Results show that assigning vehicles to specific routes (i.e. avoiding that all vehicles drive on all routes) reduces travelled distances by empty vehicles and improves the overall system performance. If all vehicles are allowed to drive on all the routes, then a certain level of randomness in agent's choice is found to be beneficial for the system performance.

The first results show that the model is able to reproduce different system configurations and to monitor, via appropriate indicators, its performance.

Further research will focus on: (a) comparing DRST with pure taxi and pure bus services; (b) testing other strategies to optimize the service (i.e. increase load factor, reduce vehicle-km), e.g. re-balancing/idle strategies; (c) testing reactive/adaptive agent behaviors for route choice strategies based on system states; (d) testing pricing strategies and public subsidies to increase the service effectiveness (in terms of satisfied demand); (e) testing the performance of the system with autonomous vehicles; (f) including elasticity of demand to price; (g) improving the demand model (e.g. including socio-demographics characteristics, data from surveys).

The final aim is to have a reliable decision-support tool for planning, management and optimization of DRST services, which can help to reduce the burden of transport in our cities and contribute to sustainable mobility.